\documentclass[12pt]{article}

\pdfoutput=1
\usepackage[latin9]{inputenc}
\usepackage[top=80pt,bottom=85pt,left=67pt,right=67pt]{geometry}
\usepackage{amssymb,amsmath}
\usepackage{xypic,subfigure}
\usepackage{graphicx,color}
\usepackage{bbm}
\usepackage{float}
\usepackage{cite}
\usepackage[debug,pageanchor=false]{hyperref}
\definecolor{link}{rgb}{.8,.15,.1}
\hypersetup{colorlinks=true,linkcolor=link,citecolor=link,urlcolor=link,linktocpage}

\usepackage{array}
\usepackage{color}
\usepackage{colortbl}

\makeatletter
\@addtoreset{equation}{section}
\makeatother

\newcommand{\beq}{\begin{equation}}
\newcommand{\eeq}{\end{equation}}
\newcommand{\bea}{\begin{eqnarray}}
\newcommand{\eea}{\end{eqnarray}}
\newcommand{\nn}{\nonumber}

\newcommand{\eq}{\begin{equation}}
\newcommand{\feq}{\end{equation}}
\newcommand{\eqn}{\begin{eqnarray}}
\newcommand{\feqn}{\end{eqnarray}}

\begin{document}
\begin{titlepage}

\begin{center}

\vskip .5in 
\noindent

{\Large \bf{AdS$_3$ T-duality and evidence for ${\cal N}=5,6$ superconformal quantum mechanics }}

\bigskip\medskip

Andrea Conti$^{a,b,}$\footnote{contiandrea@uniovi.es}  \\

\bigskip\medskip
{\small 

$a$: Department of Physics, University of Oviedo, Avenida. Federico Garcia Lorca s/n, 33007 Oviedo}

\medskip
{\small and}

\medskip
{\small 

Instituto Universitario de Ciencias y Tecnolog\'ias Espaciales de Asturias (ICTEA),\\
Calle de la Independencia 13, 33004 Oviedo, Spain}

\bigskip\bigskip
{\small

	$b$: SISSA International School for Advanced Studies, Via Bonomea 265, 34136 Trieste
	
\medskip
{\small and}

INFN - Sezione di Trieste, Via Valerio 2, I-34127 Trieste, Italy}

\vskip 2cm 

     	{\bf Abstract }
     	\end{center}
     	\noindent
	We construct two families of AdS$_2$ vacua in Type IIB Supergravity performing U(1) and SL(2) T-dualities on the $\text{AdS}_3 \times \text{$ \widehat{\mathbb{CP}}\!\!~^3$} \times $ I solutions to Type IIA recently reported in \cite{Macpherson:2023cbl}. Depending on the T-duality we operate, we find two different classes of solutions of the type $\text{AdS}_2 \times \text{$ \widehat{\mathbb{CP}}\!\!~^3$} \times $ I $\times$ I and $\text{AdS}_3 \times \text{$ \widehat{\mathbb{CP}}\!\!~^3$} \times $ I  $\times$ S$^1$. This provides evidence for more general classes of solutions  $\text{AdS}_2 \times \text{$ \widehat{\mathbb{CP}}\!\!~^3$} \times \Sigma $, dual to superconformal quantum mechanics with ${\cal N}=5,6$ supersymmetry.

\noindent
 
\vfill
\eject

\end{titlepage}

\tableofcontents

\newpage

\section{Introduction}

Conformal field theories in two dimensions (CFT$_2$) play an important role in theoretical physics in diverse areas such as, string theory, black-hole physics, condensed matter and quantum information theory. Through the AdS/CFT correspondence CFT$_2$s gain a dual description in terms of solutions of supergravity containing an AdS$_3$ factor, describing the near horizon geometries of black strings. In recent years this has played a big role towards understanding the microscopic description of these objects. Similarly, the near horizon limit of black hole geometries exhibits an AdS$_2$  factor, and thanks to the AdS/CFT correspondence we can identify their microscopical degrees of freedom within dual superconformal quantum mechanics (SCQM). The best understood avatars of the correspondence are those that preserve some degree of supersymmetry, hence, the construction of AdS$_2$ and AdS$_3$ geometries and their associate dual superconformal theories has attracted a lot of interest. Particular promising results have been obtained for the class of black holes and black strings with  ${\cal N}=4$ and $ {\cal N} = (0, 4) $ supersymmetries, such as \cite{Couzens:2017way,Couzens:2017nnr,Lozano:2019jza,Lozano:2019zvg,Lozano:2019ywa,Passias:2019rga,Lozano:2020bxo,Faedo:2020nol,Lozano:2020txg,Lozano:2020sae,Passias:2020ubv,Faedo:2020lyw,Lozano:2021rmk,Ramirez:2021tkd,Lozano:2021fkk,Couzens:2021veb}. For a specific class of these configurations, the explicit ${\cal N}=4$ superconformal quantum mechanics has been derived from a microscopic descrition \cite{Anninos:2013nra,Mirfendereski:2020rrk}.\\
~\\
Another application of low dimensional AdS spaces is in the context of defect conformal field theories \cite{Karch:2000gx,DeWolfe:2001pq,Aharony:2003qf,DHoker:2006vfr,Lunin:2007ab}. Geometries dual to conformal defects in CFT$_d$ contain warped AdS$_p$ factors, with $p<d$ at generic points in the space and are asymptotic to AdS$_{d+1}$ at isolated points. In the simplest cases, given an existing brane intersection giving rise to AdS in the near horizon, one expects to be able to realise a dual to a conformal defect by adding additional branes to the original intersection and again taking the near horizon limit. With the goal to construct dual to defect in mind, a lot of attempts at finding classifications of AdS$_3$ and AdS$_2$ solutions together with their dual CFTs with different amount of supersymmetry have been carried out and remarkable progress has been achieved. Notable examples are \cite{Lozano:2022vsv,Dibitetto:2017klx,Dibitetto:2017tve,Dibitetto:2018gtk,Dibitetto:2020bsh,Chen:2020mtv,Chiodaroli:2009xh,Chiodaroli:2009yw,Dibitetto:2018iar,Hong:2019wyi,Dibitetto:2019nyz}.\\
~\\
In general, there are two different tools we can use to construct a solution with AdS$_2$ from AdS$_3$ and vice versa. First of all, if the solution contains an S$^2$ or an S$^3$, we can perform a double analytic continuation on them and the AdS metric: AdS$_3$ $\times$ S$^2$ becomes AdS$_2$ $\times$ S$^3$ and vice versa. From the string theory point of view there is another interesting property: we can connect AdS$_3$ and AdS$_2$ via T-duality. Both these ideas were put to work in \cite{Conti:2023naw}. \\
~\\
This paper follows these ideas. In \cite{Conti:2023naw} the general procedure for generating AdS$_2$ solutions from existing AdS$_3$ ones using U(1) and SL(2) T-dualities was derived. It was further established that when the AdS$_3$ solutions preserve some amount of chiral supersymmetry, it is possible to arrange for the AdS$_2$ solution to preserve all of these, this is true only in the case the supercharges have the same chirality. Specifically, if the AdS$_3$ solution preserves two types of chiral supersymmetry ${\cal{N}}=(p,q)$ we can arrange the AdS$_2$ solution to preserve ${\cal{N}}=p$ or ${\cal{N}}=q$ but not ${\cal{N}}=p+q$ \footnote{Similar results were proven for the SU(2)-duality, in \cite{Kelekci:2014ima} it was shown that the SUSY variations were mapped up to the Kosmann derivative for SU(2) transformations of S$^3$. }. Depending on the T-duality procedure we wanted to carry out, i.e. U(1) or SL(2) T-duality, we found two different types of solutions. \\
~\\
In this work we will apply this procedure to two classes of AdS$_3$ solutions constructed in \cite{Macpherson:2023cbl}, which preserve ${\cal N}=(5,0)$ or ${\cal N}=(6,0)$ supersymmetry and realise the superconformal algebras $\mathfrak{osp}(5|2)$ and $\mathfrak{osp}(6|2)$. This results in four new classes of AdS$_2$ solutions, realising these superconformal algebras, as summarized in Table \ref{table 1}. 

\begin{table}[h!]
\label{table 1}
\centering
\begin{tabular}{|c|c|c|c|}
\hline
SUSY & Type IIA & Type IIB: SL(2) T-Duality & Type IIB: U(1) T-Duality \\
\hline
${\cal N}=(6,0)$ & AdS$_3$ $\times$ $\mathbb{CP}^3$ $\times$ I & AdS$_2$ $\times$ $\mathbb{CP}^3$ $\times$ I $\times$ I  & AdS$_2$ $\times$ $\mathbb{CP}^3$ $\times$ I $\times$ S$^1$ \\
\hline
${\cal N}=(5,0)$ & AdS$_3$ $\times$ $\widehat{\mathbb{CP}}\!\!~^3$ $\times$ I & AdS$_2$ $\times$ $\widehat{\mathbb{CP}}\!\!~^3$ $\times$ I $\times$ I  & AdS$_2$ $\times$ $\widehat{\mathbb{CP}}\!\!~^3$ $\times$ I $\times$ S$^1$ \\
\hline
\end{tabular}
\caption{Summary of the seed solutions and their duals}
\end{table}

\noindent This provides strong evidence that more general classes of AdS$_2$ $\times$ $ \widehat{\mathbb{CP}}\!\!~^3$ $\times$ $\Sigma$ solutions with these supersymmetries and $\Sigma$ a $2d$ Riemann surface should exist. Similar constructions have been studied in the past \cite{Lozano:2020txg,Lozano:2021fkk,Chen:2020mtv,DHoker:2007mci}. This generalization could be interesting to explore, along with the identification of their dual SCQMs.


\section{AdS$_3$ $\times$ $ \widehat{\mathbb{CP}}\!\!~^3$ $\times$ I} \label{sect:2}
In this section we briefly summarize the construction of AdS$_3$ vacua in Type IIA supergravity carried out in \cite{Macpherson:2023cbl}. We start presenting the general class of solutions, to then turn to the two cases with $\mathfrak{osp}(n|2)$ superconformal algebra with $n=5, 6$. \\
~\\
The general decomposition in Type II Supergravity associated to a vacuum solution with a three dimensional Anti de Sitter spacetime, is a warped product of AdS$_3$ with M$_7$
\begin{align}
ds^2 & = e^{2A} ds^2(\text{AdS}_3)+ ds^2(\text{M}_7), \nn \\
H & = e^{3A}c_0\text{vol}(\text{AdS}_3) + H_3,~~~ F=f_{\pm}+e^{3A}\text{vol}(\text{AdS}_3)\star_7\lambda(f). \label{eq:AdS3vacua}
\end{align}
Generally M$_7$ can be any seven dimensional manifold. For the solutions in \cite{Conti:2023naw}, M$_7$ is a foliation of a (squashed) complex projective 3 space ($\mathbb{CP}\!\!~^3$) over an interval. The resulting isometries are the SO(2,2) group associated to the AdS$_3$ metric and the SO(5) or SO(6) isometry groups of $\widehat{\mathbb{CP}}\!\!~^3$ or $\mathbb{CP}^3$, where by $\widehat{\mathbb{CP}}\!\!~^3$ we denote a squashed $\mathbb{CP}^3$ preserving precisely SO(5) isometries.
$(e^{2A},f,H_3)$ in \eqref{eq:AdS3vacua} and the dilaton $\Phi$ must have support on M$_7$ so as to preserve the SO(2,2) symmmetry of AdS$_3$. $H$ is the NS three form, which we choose to be purely magnetic\footnote{Otherwise it would not be possible to operate a SL(2) T-duality since $H$ would not be invariant under this group.}. The RR fluxes $F$ are encoded as a polyform of even/odd degree in IIA/IIB. The polyform must satisfy the self-duality constraint
\begin{equation}
F= \star \lambda (F), \nn
\end{equation}
where the function $\lambda$ acts on a p-form as $\lambda(X_p)= (-1)^{[\frac{p}{2}]}X_p$. The metric in \cite{Macpherson:2023cbl} reads
\begin{align}
ds^2(\text{M}_7)&= e^{2 K}dr^2+ ds^2(\widehat{\mathbb{CP}}\!\!~^3)\label{eq:defCP3squahsed}\\[2mm]
ds^2(\widehat{\mathbb{CP}}\!\!~^3)&= \frac{1}{4}\bigg[e^{2C} \left(d\alpha^2+ \frac{1}{4}\sin^2\alpha (L_i)^2\right)+ e^{2D} (Dy_i)^2\bigg],~~~~~ Dy_i= dy_i+\cos^2\left(\frac{\alpha}{2}\right)\epsilon_{ijk} y_j L_k,\nn
\end{align}
where $r$ is the coordinate along the interval, $(e^{2A},e^{2C},e^{2D},e^{2 K},\Phi)$ depend only on $r$, $L_i$ are the left invariant forms on SU(2) and $y_i$ are embedding coordinates on the unit radius 2-sphere. If $e^{2C}$ and $e^{2D}$ are equal, a round $\mathbb{CP}^3$ is restored. For more details about the geometry of the solution, we refer to \cite{Macpherson:2023cbl}.
In the article just mentioned, a general solution has been built: it effectively preserves a $\mathfrak{osp}(n|2)$ superalgebra on the conformal field theory side, where $n$ takes values of either $5$ or $6$. By appropriately fixing the parameter to a specific value, the authors were able to obtain either $\mathfrak{osp}(6|2)$ or $\mathfrak{osp}(5|2)$. Interestingly, the case of $\mathfrak{osp}(6|2)$ corresponds to a round $\mathbb{CP}^3$ manifold. On the other hand $\mathfrak{osp}(5|2)$ leads to an SO(5) preserving squashing of $\mathbb{CP}^3$, which we denote as $ \widehat{\mathbb{CP}}\!\!~^3$. We present these classes of solutions in the next section.

\subsection{Type IIA: general class} \label{sec:general class}
The general class of AdS$_3$ solutions realising ${\cal N}=(n,0)$ for $n=5,6$ has NS sector given by
\begin{align}
\label{eq:3.0}
\frac{ds^2}{2 \pi} & = \frac{h u}{\sqrt{\Delta_1}} ds^2(\text{AdS}_3) +\frac{\sqrt{\Delta_1}}{4 u}\left[   \frac{2}{ h''} \left( ds^2(\text{S}^4)+\frac{1}{\Delta_2}(Dy_i)^2 \right) + \frac{1}{h} dr^2 \right], \nn \\[2mm]
e^{-\Phi} & =\frac{h''^{\frac{3}{2}} \sqrt{u} \sqrt{\Delta_1} }{2 \sqrt{\pi } \Delta_2^{\frac{1}{4}}}, \qquad \Delta_1=2 h u^2 h''-\left(u h'-h u'\right)^2, \qquad \Delta_2=1 + \frac{2 h' u'}{u h''}, \qquad H_3=dB_2, \nn \\[2mm]
B_2 & = 4\pi \left[- \left((r-k)-\frac{uh'-u'h}{uh''}\right) J_2+ \frac{u'}{2h''}\left( \frac{h}{u} + \frac{hh''-2(h')^2}{2h'u'+uh''}\right)(J_2-\tilde{J}_2) \right],
\end{align}
where $u$ and $h$ are functions of $r$ and $k$ is a constant (we choose $u$, $h$ and $h''$ to be positive). The RR sector is given by
\begin{align}
\label{eq:3.1}
F_0 & = -\frac{h'''}{2 \pi}, \qquad F_2 = B_2F_0+2(h''-(r-k)h''')J_2, \nn \\[2mm]
F_4 & = \pi d\left(h'+ \frac{hh''u(uh'+u'h)}{\Delta_1} \right)\wedge  \text{vol}(\text{AdS}_3)+ B_2 \wedge F_2 - \frac{1}{2}B_2\wedge B_2 F_0 \nn \\[2mm]
& -4 \pi \left[ (2h'+(r-k)(-2h''+(r-k)h''')J_2 \wedge J_2+d\left(\frac{hu'}{u}\right) \text{Im}\Omega_3 \right],
\end{align}
The solution is defined in terms of two ODES: first
\begin{align}
u''=0,
\end{align}
which is required to hold globally by supersymmetry. Second, the Bianchi identities of the fluxes impose
\begin{align}
h'''=0,
\end{align}
in regular parts of a solution. $J_2$ is an SO(6) invariant Kahler form such that
\begin{align}
J_2 \wedge J_2 \wedge J_2 = 6 \text{vol}(\mathbb{CP}^3), \nn
\end{align}
and $\tilde{J}_2$ and $\Omega_3$ are a real 2-form  and a complex 3-form respectively that are invariant under an SO(5) subgroup of the whole isometry group of $\mathbb{CP}^3$. They obey the following identities
\begin{align}
&J_2\wedge J_2\wedge J_2=\tilde J_2\wedge \tilde J_2\wedge \tilde J_2=\frac{3 i}{4}\Omega_3\wedge\overline{\Omega}_3,~~~~J_2\wedge J_2+ \tilde J_2\wedge \tilde J_2=2 \tilde J_2\wedge J_2,\nn\\[2mm]
&J_2\wedge \Omega_3=\tilde{J}_2\wedge \Omega_3=0,~~~~ d J_2=0,~~~~ d\tilde J_2= 4 \text{Re}\Omega_3,~~~~d \text{Im}\Omega_3=6 \tilde J_2\wedge J_2-2 J_2\wedge J_2,
\end{align}
For more details we refer to \cite{Macpherson:2023cbl}. \\
~\\
In the following subsections we distinguish between two different classes of solutions: $\mathbb{CP}\!\!~^3$ and $\widehat{\mathbb{CP}}\!\!~^3$. The first one is realised when $u=$ constant, the second one when $u \neq$ constant. However, if $u=$ constant, then it actually drops out of the class of solutions, so it is sufficient to fix $u=1$. For this tuning of $u$, the round $\mathbb{CP}\!\!~^3$ is restored and ${\cal{N}}=(6,0)$ supersymmetry is preserved. This solution is presented in \ref{sec:AdS3 osp62}. If $u' \neq 0$ it is possible to use diffeomorphism invariance to fix $u=r$ as shown in \cite{Macpherson:2023cbl}. For this tuning of $u$, $\widehat{\mathbb{CP}}\!\!~^3$ is squashed such that only its SO(5) subgroup is preserved. Solutions of this type preserve ${\cal{N}}=(5,0)$ supersymmetry. We present this class in \ref{sec:AdS3 osp52}. \\
~\\
Depending on how we tune $h$, the domain of $r$ can be infinite, semi-infinite or bounded at both ends. When the bounds are present, it is possible to arrange for them to be physical singularities. Depending on the amount of supersymmetry preserved we have different cases. In ${\cal{N}}=(6,0)$ ($u=1$), the solution can be bounded by O2 planes, D8/O8 systems and a KK monopole. For ${\cal{N}}=(5,0)$ ($u=r$), the possibilities are O6, O4, O2 planes, D6-branes and again a KK monopole \cite{Macpherson:2023cbl}.

\subsubsection{AdS$_3$ $\times$ $\mathbb{CP}^3$ $\times$ I ~ ${\cal N}=(6,0)$} \label{sec:AdS3 osp62}

Specialising the previous class to the case $u=1$ we recover a round $\mathbb{CP}^3$, with SO(6) isometry group and ${\cal N}=(6,0)$ supersymmetry. The NS sector reads
\begin{align}
\frac{ds^2}{2\pi}&= \frac{h}{\sqrt{2 h h'' -( h')^2}}ds^2(\text{AdS}_3)+\sqrt{2 h h'' -( h')^2}\bigg[\frac{1}{4 h }dr^2+ \frac{2}{ h''} ds^2(\mathbb{CP}^3)\bigg],\nn\\[2mm]
e^{-\Phi}&=\frac{ (h'')^{\frac{3}{2}}}{2\sqrt{\pi}(2 h h'' -( h')^2)^{\frac{1}{4}}},~~~~
H=dB_2,~~~~ B_2= 4\pi\left(-(r-k)+\frac{ h'}{ h''}\right) J_2,\label{eq:neq6nssector}
\end{align}
and the RR sector
\begin{align}
F_0&=-\frac{1}{2\pi}h''',~~~~~F_2= B_2 F_0+  2(h''-(r-k) h''')J_2,\nn\\[2mm]
F_4&=\pi d\left(h'+\frac{hh'h''}{2 h h'' -( h')^2}\right)\wedge \text{vol}(\text{AdS}_3)+B_2 \wedge F_2-\frac{1}{2}B_2\wedge B_2 F_0\nn\\[2mm]
& -4\pi(2h'+(r-k)(-2 h''+(r-k) h'''))J_2\wedge J_2.
\end{align}

\subsubsection{AdS$_3$ $\times$ $ \widehat{\mathbb{CP}}\!\!~^3 $ $\times$ I ~ ${\cal N}=(5,0)$}   \label{sec:AdS3 osp52}
Setting $u' \neq 0$ we recover the general class with ${\cal N}=(5,0)$ supersymmetry. In particular we can fix $u=r$ without loss of generality. The resulting NS sector takes the form
\begin{align}
\frac{ds^2}{2\pi}&= \frac{hr}{\sqrt{\Delta_1}}ds^2(\text{AdS}_3)+\frac{\sqrt{\Delta_1}}{4r}\bigg[ \frac{2}{ h''} \bigg(ds^2(\text{S}^4)+ \frac{1}{\Delta_2 }(Dy_i)^2\bigg)+\frac{1}{h }dr^2\bigg], \nn \\[2mm]
e^{-\Phi}&=\frac{ h''^{\frac{3}{2}}\sqrt{r}\sqrt{\Delta_2 }}{2\sqrt{\pi}\Delta_1^{\frac{1}{4}}},~~~~\Delta_1=2 h h'' r^2-(r h'- h )^2,~~~~\Delta_2=1+\frac{2 h' }{r h''},~~~~H=dB_2, \nn \\[2mm]
 B_2&= 4\pi\bigg[\left(-(r-k)+\frac{r h'-h }{r h''}\right) J_2+ \frac{1}{2h''}\left(\frac{h}{r}+\frac{h h''-2 (h')^2}{2h'+r h''}\right)\left(J_2- \tilde J_2\right)\bigg],\label{eq:summerystart}
\end{align}
while the $d=10$ RR fluxes are then  given by
\begin{align}
F_0&=-\frac{1}{2\pi}h''',~~~~~F_2= B_2 F_0+  2(h''-(r-k) h''')J_2,\nn \\[2mm]
F_4&=\pi d\left(h'+\frac{hh''r(rh'+ h )}{\Delta_1}\right)\wedge \text{vol}(\text{AdS}_3)+B_2 \wedge F_2-\frac{1}{2}B_2\wedge B_2 F_0\nn \\[2mm]
& -4\pi\bigg[(2h'+(r-k)(-2 h''+(r-k) h'''))J_2\wedge J_2+ d\left(\frac{h }{r}\text{Im}\Omega\right)\bigg].
\end{align}

\section{Type IIB: T-dual solutions }

In this section we present the T-dual solutions in Type IIB Supergravity constructed via U(1) and SL(2) T-dualities as explained in \cite{Conti:2023naw}. We start presenting the T-dual general solutions, i.e. the one from which we can derive the two specfic cases with ${\cal N}=n$  for $n=5,6$. Recalling the results obtained in \cite{Conti:2023naw}, since we start from a class of solutions preserving chiral supersymmetry, the T-dual solutions preserve all of this (although the supersymmetry should no longer be viewed as chiral). 
We denoted $\rho$ the T-dual coordinate, in both the Abelian and Non-Abelian cases. 

\subsection{General class AdS$_2$ $\times$ $ \widehat{\mathbb{CP}}\!\!~^3$ $\times$ I $\times$ I}  \label{NATD general}
Operating an SL(2) T-duality along AdS$_3$, we break this spacetime into an AdS$_2$ factor and a semi-infinite interval spanned by $\rho$. The bounds of $r$ are discussed in Section \ref{sect:2} and remain the same for the T-dual solution. The interval of $\rho$ is semi-infinite because the warp factor of AdS$_2$ blows up when $\rho^2 =  \frac{\pi^2 h^2 u^2}{\Delta_1}$, we can without loss of generality take $\rho \in \left[\frac{\pi h u}{\Delta_1},0\right)$ but the lower bound is a function of $r$. Typically, at regular points along $r$, the lower bound gives OF1-planes \cite{Ramirez:2021tkd}. At singular points along $r$, things are more complicated and this depends on the type of singularity that is present, we will not attempt to perform a detailed analysis here. \\
The T-dual NS sector is given by
\begin{align}
\label{NATD general NS}
d\hat{s}^2 &=  \frac{ \pi  \rho^2  hu \sqrt{\Delta_1}}{2 \left( \rho^2\Delta_1-\pi ^2 h^2 u^2\right)} ds^2(\text{AdS}_2) + \frac{\pi \sqrt{\Delta_1}}{ u h''} \left[ ds^2(\text{S}^4)+\frac{1}{\Delta_2}(Dy_i)^2  \right] +\frac{\pi \sqrt{\Delta_1}}{2 h u} \left[dr^2 + \frac{d\rho^2}{\pi^2} \right], \nn \\[2mm]
e^{-\hat{\Phi}} & =\frac{\sqrt{h u} h'' \sqrt{u h''+2 h' u'}\sqrt{ \rho^2-\frac{\pi^2 h^2 u^2}{\Delta_1 }}}{ \sqrt{2} \sqrt{\Delta_1 }}, ~~~ \tilde{H}= d\tilde{B}, ~~~ \tilde{B} = B_2 + \frac{\rho^3}{2 \left(\rho^2-\frac{\pi^2 h^2 u^2}{\Delta_1 }\right)} \text{vol}(\text{AdS}_2).
\end{align}
As emphasized in the previous section, $u$, $h$ and $h''$ are positive. The Bianchi identity sets $h''''=0$, with possible $\delta$-function sources at the loci of D8 branes. We write the RR sector in terms of the Page fluxes defined as
\begin{align}
\hat{F}=e^{-\tilde{B}} \wedge F. \nn 
\end{align}
They are known to be closed away from sources, $d\hat{F}=0$. They are
\begin{align}
\label{NATD general RR}
\hat{F}_1 & = \frac{ \rho h'''}{2 \pi} d\rho + p_1(r) dr, \nn \\[2mm]
\hat{F}_3 & = \frac{\rho^2 h''' }{4 \pi } d\rho \wedge \text{vol}(\text{AdS}_2) + 2 \pi^2 d\left( \frac{\Delta_s h^2 u'}{\Delta_1 } \tilde{J}_2\right)-2 \rho \left(h''-(r-k) h'''\right) d\rho\wedge J_2 + J_2 \wedge d(p_2(r)), \nn  \\[2mm]
\hat{F}_5 & = d(( p_3(r,\rho) d\rho + p_4(r) dr)\wedge \text{Im}\Omega_3 ) +d(q(r,\rho))\wedge J_2 \wedge J_2 \nn \\[2mm]
& - d\left(\frac{1}{3} \rho^3\left( h''- (r-k) h'''\right)\right)\wedge \text{vol}(\text{AdS}_2) \wedge J_2.
\end{align}
The expressions for $p_i$ and $Q$ are given in Appendix \ref{subsection:Appendix 1 A}.

\subsubsection{AdS$_2$ $\times$ $\mathbb{CP}^3$ $\times$ I $\times$ I ~ ${\cal N}=6 $ } \label{NATD n=6}
Here we present the class of solutions that preserves ${\cal N} = 6$ supersymmetry. We set $u=1$ in \eqref{NATD general NS} and \eqref{NATD general RR}, in this way the round $\mathbb{CP}^3$ is restored. The NS sector is given by
\begin{align}
d\hat{s}^2&= \frac{ \pi  \rho^2 h \sqrt{2 h h''-h'^2}}{2 \left(\rho^2 \left(2 h h''-h'^2\right)-\pi ^2 h^2\right)}ds^2(\text{AdS}_2) + \frac{ 4 \pi \sqrt{2 h h'' -( h')^2}}{h''} ds^2(\mathbb{CP}^3) \nn \\[2mm]
& + \frac{ \pi \sqrt{2 h h'' -( h')^2}}{2 h } \left[ dr^2 + \frac{d\rho^2}{\pi^2}  \right] , ~~~~ \tilde{H} = d\tilde{B}, \\[2mm]
e^{-\Phi}&= \frac{\rho h''^{\frac{3}{2}} \sqrt{\frac{\pi^2 h^3}{\rho^2 \left(h'^2-2 h h''\right)}+h}}{\sqrt{4 h h''-2 h'^2}},~~~~
\tilde{B} = B_2 + \frac{\rho^3 \left(h'^2-2 h h''\right)}{2  \left( \pi ^2 h^2 + \rho^2( h'^2 -2 h h'' ) \right)} \text{vol}(\text{AdS}_2), \nn
\label{eq:neq6nssector}
\end{align}
where $B_2$ is defined in \eqref{eq:neq6nssector}, and the Page fluxes are
\begin{align}
\hat{F}_1&= \frac{ \rho h'''}{2 \pi } d \rho +\frac{\pi  h \left(3 h''^2 \left(2 h h''-h'^2\right)-h''' h'^3\right)}{2 \left(2 h h''-h'^2\right)^2} dr, \nn \\[2mm]
\hat{F}_3&= \frac{\rho^2 h''' }{4 \pi }\text{vol}(\text{AdS}_2)\wedge d\rho -2 \rho \left(h''-(r-k) h'''\right) d\rho\wedge J_2, \nn \\[2mm]
& + J_2 \wedge d\left(-2 \pi ^2 \left(\left(2 h-(r-k) h'\right)+\frac{h h' \left(h'-(r-k) h''\right)}{2 h h''-h'^2}\right)\right) \nn \\[2mm]
\hat{F}_5&= d(a(r,\rho))\wedge J_2\wedge J_2-d\left(\frac{1}{3} \rho^3 \left(h''-(r-k) h'''\right)\right)\wedge \text{vol}(\text{AdS}_2)\wedge J_2,
\end{align}
the functions $a(r,\rho)$ and $b'(r)$ are given by 
\begin{align}
a(r,\rho)& = b(r) + 2 \pi  \rho^2 \left(2 h'+(r-k) \left(-2 h''+(r-k) h'''\right)\right), \nn \\[2mm]
b'(r) & = \frac{\left(4 \pi ^3 h\right) }{3 \left(2 h h''-h'^2\right)^2} \bigg(6 \left(h'^2 \left(h h''-h'^2\right)+2 h^2 \left(h''^2-h''' h'\right)\right) \nn \\[2mm]
& -3 (r-k)^2 \left(-6 h h''^3+h''' h'^3+3 h'^2 h''^2\right) -6 (r-k) h' \left(6 h h''^2-2 h h''' h'-3 h'^2 h''\right)\bigg), \nn 
\end{align}
where $k$ is a constant.

\subsubsection{AdS$_2$ $\times$ $\widehat{\mathbb{CP}}\!\!~^3$ $\times$ I $\times$ I ~ ${\cal N}=5$ } \label{NATD n=5}

To construct the solutions that preserve ${\cal N}=5$ we must set $u=r$. If we do this, we find
\begin{align}
d\hat{s}^2 &=  \frac{ \pi  \rho^2  hr \sqrt{\Delta_1}}{2 \left( \rho^2\Delta_1-\pi ^2 h^2 r^2\right)} ds^2(\text{AdS}_2) +  \frac{ \pi \sqrt{\Delta_1}}{ r h''}\left[ ds^2(\text{S}^4)+\frac{1}{\Delta_2}(Dy_i)^2  \right] +  \frac{ \pi \sqrt{\Delta_1}}{2 r h} \left[ dr^2  + \frac{ d\rho^2 }{\pi^2} \right],  \nn \\[2mm]
e^{-\hat{\Phi}} & =\frac{\sqrt{h r} h'' \sqrt{r h''+2 h'}\sqrt{ \rho^2-\frac{\pi^2 h^2 r^2}{\Delta_1 }}}{ \sqrt{2} \sqrt{\Delta_1 }}, ~~~~~ \tilde{H} = d\tilde{B}, \nn \\[2mm]
\tilde{B} & = B_2 + \frac{\rho^3}{2 \left(\rho^2-\frac{\pi^2 h^2 r^2}{\Delta_1 }\right)} \text{vol}(\text{AdS}_2).
\end{align}
$\Delta_1$ and $\Delta_2$ are defined in \eqref{eq:3.0} with $u=r$. About the RR sector, the Page fluxes are the same as the ones reported in \eqref{NATD general RR} and the polynomials defined in \eqref{eq NATD poly} with the specified
\begin{align}
\Delta_s = h-r h',~~~~~ \Delta_r = h-r h'.
\end{align}

\subsection{General class AdS$_2$ $\times$ $\widehat{\mathbb{CP}}\!\!~^3$ $\times$ I $\times$ S$^1$} \label{ATD general}
In this section we present the solutions in Type IIB Supergravity via U(1) T-duality. In this case the spacetime we obtain is a compact one with a S$^1$ replacing the semi-infinite interval. For more details we refer to \cite{Conti:2023naw}. We denote the T-dual coordinate by $\rho$ again. The domain of $r$ remains the same as it was in Type IIA where it is possible to bound $r$ by physical singularities. The nature of these sources changes under the duality. Two situations can happen depending on whether we T-dualize on an isometry parallel or orthogonal to the source. In the former case the source loses a worldvolume direction and gains a co-dimension over which it is smeared. In the latter case the source gains a worldvolume direction, i.e. a Dp-brane gets mapped to a D(p$+1$)-brane if we T-dualize along one of its co-dimensions while it becomes a D(p$-1$)-brane smeared over the dualisation isometry if we T-dualize within its worldvolume. Formally, the story is analogous for Op-planes\footnote{Up to the well known subtleties in the relation with the smearing of Op-planes.}. \\
The NS sector is
\begin{align}
\label{ATD general NS}
d\hat{s}^2 &= \frac{\pi hu}{2 \sqrt{\Delta_1}} ds^2(\text{AdS}_2) +\frac{\pi \sqrt{\Delta_1}}{ u h''} \left[ ds^2(\text{S}^4)+\frac{1}{\Delta_2}(Dy_i)^2  \right] + \frac{\pi \sqrt{\Delta_1}}{ 2 h u} \left[ dr^2 +  \frac{d\rho^2}{\pi^2} \right] , \nn \\[2mm]
\tilde{B} & = B_2 + \frac{ \rho}{2}\text{vol}(\text{AdS}_2) , ~~~~~ \tilde{H} = d \tilde{B}, ~~~~~ e^{-\hat{\Phi}} = \frac{\sqrt{h} \sqrt{u} h'' \sqrt{u h''+2 h'u'}}{\sqrt{2}\sqrt{\Delta_1}}.
\end{align}
where $B_2$ is given by \eqref{eq:3.0}. 
We write the RR sector using the Page fluxes like in the previous section
\begin{align}
\label{ATD general RR}
\hat{F}_1 & = -\frac{ h'''}{2 \pi } d\rho, ~~~~~  \hat{F}_3 = p_5(r,\rho) \text{vol}(\text{AdS}_2) \wedge dr + d(p_6(r,\rho))\wedge J_2 + \frac{\rho h'''}{4 \pi }  \text{vol}(\text{AdS}_2) \wedge d \rho, \nn \\[2mm]
\hat{F}_5 & = p_7(r) d\rho \wedge dr \wedge \text{Im}\Omega_3 + p_8(r) d\rho \wedge J_2 \wedge J_2 + d(Q(r))\wedge \text{vol}(\text{AdS}_2) \wedge J_2 \nn \\[2mm]
& + p_9(r) d\rho \wedge \tilde{J}_2 \wedge \tilde{J}_2 + \frac{1}{4} d(p_{10}(r))\wedge \text{vol}(\text{AdS}_2) \wedge \tilde{J}_2+ p_{10}(r) \text{vol}(\text{AdS}_2) \wedge \text{Re}\Omega_3 \nn \\[2mm]
&- \rho \left(h'' - (r-k) h'''\right) \text{vol}(\text{AdS}_2) \wedge d \rho \wedge J_2 .
\end{align}
The specific expressions for the functions appearing here are given in Appendix \ref{subsection:Appendix 1 B}.

\subsubsection{AdS$_2$ $\times$ $\mathbb{CP}^3$ $\times$ I $\times$ S$^1$ ~ ${\cal N}=6$ } \label{ATD n=6}
As in the case of the T-duality via SL(2), we can differentiate two solutions according to the value we set for $u$, preserving ${\cal N}=5$ or $6$ supersymmetry. In this section we present the solution with ${\cal N}=6$, obtained by fixing $u=1$, that reproduces the round metric on $\mathbb{CP}^3$. The NS sector reads
\begin{align}
d\hat{s}^2 & = \frac{\pi  h}{2 \sqrt{2 h h''-h'^2}} ds^2(\text{AdS}_2) +  \frac{4 \pi}{h''} \sqrt{2 h h'' -( h')^2}ds^2(\mathbb{CP}^3) + \frac{ \pi \sqrt{2 h h'' -( h')^2}}{2 h }\left[dr^2 + \frac{d\rho^2}{\pi^2}\right]  , \nn \\[2mm]
\tilde{B} & = B_2 + \frac{ \rho}{2} \text{vol}(\text{AdS}_2) , ~~~~~ \tilde{H} = d\tilde{B}, ~~~~~  e^{-\hat{\Phi}} = \frac{\sqrt{h} h''^{\frac{3}{2}}}{\sqrt{4 h h''-2 h'^2}}.
\end{align}
As in the previous case, we write the RR sector using the Page fluxes
\begin{align}
\hat{F}_1 & = -\frac{ h'''}{2 \pi } d\rho, \nn \\[2mm]
\hat{F}_3 & =  d\left(2 \rho \left(h'' - (r-k) h''' \right)\right)\wedge J_2 - \frac{\pi  h \left(h''' h'^3+ 3 \left(h'^2 - 2 h h''\right) \right)}{4 \left(2 h h''-h'^2\right)^2} \text{vol}(\text{AdS}_2) \wedge dr  \nn \\[2mm]
& + \frac{\rho h'''}{4 \pi }  \text{vol}(\text{AdS}_2) \wedge d \rho , \nn \\[2mm]
\hat{F}_5 & = - 4 \pi  \left(2 h'+(r-k) \left(-2 h''+(r-k) h'''\right)\right) d\rho \wedge J_2 \wedge J_2 + d(c(r))\wedge \text{vol}(\text{AdS}_2) \wedge J_2 \nn \\[2mm]
& - \rho \left(h'' - (r-k) h'''\right) \text{vol}(\text{AdS}_2) \wedge d \rho \wedge J_2 , \nn \\[2mm]
c'(r) & = \frac{\pi^2 h \left(\left(2 h h''' h'^2+3 h'^3 h''-6 h h' h''^2 \right)+(r-k) \left(6 h h''^3-h''' h'^3-3 h'^2 h''^2 \right)\right)}{\left(2 h h''-h'^2\right)^2}.
\end{align}

\subsubsection{AdS$_2$ $\times$ $\widehat{\mathbb{CP}}\!\!~^3$ $\times$ I $\times$ S$^1$ ~ ${\cal N}=5$ } \label{ATD n=5}

For the case of ${\cal N}=5$ we can without loss of generality fix $u=r$, obtaining
\begin{align}
\label{eq:3.7}
d\hat{s}^2 &= \frac{\pi hr}{2 \sqrt{\Delta_1}} ds^2(\text{AdS}_2) + \frac{ \pi \sqrt{\Delta_1}}{ r h''}\left[ ds^2(\text{S}^4)+\frac{1}{\Delta_2}(Dy_i)^2  \right] +  \frac{\pi \sqrt{\Delta_1}}{2 h r} \left[ dr^2 + \frac{d\rho^2}{\pi^2}\right], \nn \\[2mm]
\tilde{B} & = B_2 + \frac{\rho}{2}\text{vol}(\text{AdS}_2) , ~~~~~ \tilde{H}=d\tilde{B}, ~~~~~  e^{-\hat{\Phi}} = \frac{\sqrt{h} \sqrt{r} h'' \sqrt{r h''+2 h'}}{\sqrt{2}\sqrt{\Delta_1}}.
\end{align}
Again $\Delta_1$ and $\Delta_2$ are defined in \eqref{eq:3.0} with $u=r$. For the RR sector the Page fluxes are the same ones reported in \eqref{ATD general RR} and \eqref{eq ATD poly}, with
\begin{align}
\Delta_s = h-r h',~~~~~ \Delta_r = h-r h'.
\end{align}

\section{Conclusions}

In this work we presented two new classes of AdS$_2$ $\times$ $\widehat{\mathbb{CP}}\!\!~^3$ solutions to Type IIB supergravity with ${\cal N}=6$ and ${\cal N}=5$ supersymmetry. The original AdS$_3$ solutions previous to our U(1) or SL(2) duality transformations were classified according to how much supersymmetry they preserve, ${\cal N} = (5,0)$, i.e. 10 real supercharges, or ${\cal N}=( 6,0)$, i.e. 12 real supercharges. The dual AdS$_2$ solutions generated preserve the same amount of supercharges, but supersymmetry should no longer be viewed as chiral - i.e. ${\cal N}=(n,0)$ for AdS$_3$ becomes ${\cal N}=n$ for AdS$_2$.\\
~\\
The results found in this paper may play a very interesting role in the context of the AdS/CFT correspondence, for example these solutions could be interpreted as near horizons of black holes. 
For this purpose it would be very interesting to identify the SCQM dual to our solutions. The solutions could also find a defect interpretation. Our results also provide strong evidence for the existence of more general classes of solutions with ${\cal N}=5$ and ${\cal N }=6$ supersymmetry. Indeed, there are several cases where 
T-duality has provided the first examples of much broader classes of solutions. These include supersymmetic AdS$_5$ solutions without 5-form flux, first realised in \cite{Macpherson:2014eza}, and later extended to the general class of \cite{Couzens:2016iot}, the T-duals of the D1-D5 \cite{Sfetsos:2010uq} and D4-D8 \cite{Lozano:2012au} near horizon geometries, which pointed the way to the general classes of AdS$_3$ \cite{Lozano:2019emq} and AdS$_6$ \cite{DHoker:2016ujz} solutions, etc. Solutions found in this paper strongly suggest the existence of two broader classes of AdS$_2$ $\times$ ${\cal{M}}_6 $ $\times$ $\Sigma$ solutions, with $\Sigma$ a two dimensional Riemann surface and ${\cal M}_6$ either a round or a squashed $\mathbb{CP}^3$, depending on whether ${\cal N}=6$ or ${\cal N}=5$ is preserved. These general classes of solutions should have a metric decomposing as
\begin{equation}
ds^2 = f_1^2 ds^2(\text{AdS}_2) + \frac{1}{4}\bigg[f_2^2 ds^2(\text{S}^4) + f_3^2 (Dy_i)^2\bigg] + f_4^2 ds^2(\text{$\Sigma$}),
\end{equation}
where $f_1$, $f_2$, $f_3$, $f_4$ and the dilaton are functions depending only on the coordinates on the Riemann surface. The NS three-form and the RR p-forms on the other hand will have support in $\Sigma$ and either the SO(6) or SO(5) invariant forms presented in section \ref{sec:general class}. Solving the equations of motion and the Killing spinor equations, we should find an ${\cal N}=6$ and an ${\cal N}=5$ class of solutions generalising \eqref{NATD general NS}, \eqref{NATD general RR} and \eqref{ATD general NS}, \eqref{ATD general RR} respectively. Indeed we can notice that, in our solutions, all warped factors depend only on the coordinates living on the Riemann surface spanned by $r$ and $\rho$. Our work also predicts the existence of SCQM with ${\cal N} = 6$ and ${\cal N} = 5$ supersymmetry that it would be interesting to investigate. We hope to report on these classes in the near future.

\section*{Acknowledgments}
I am thankful to Yolanda Lozano and Niall Macpherson for useful discussions and for their comments on the draft. I thank SISSA for hospitality since part of my work was done during my visiting period there. This work is supported by grants from the Spanish government MCIU-22-PID2021-123021NB-I00 and Principality of Asturias SV-PA-21-AYUD/2021/52177.

\appendix

\section{Polynomial of the general classes}
In this appendix we list the polynomials that couple in the RR sector of equations \eqref{NATD general RR} and \eqref{ATD general RR}. We start by reporting the SL(2) case, in the next section we report also the U(1) case.

\label{section:Appendix 1}

\subsection{Polynomials in AdS$_2$ $\times$ $\widehat{\mathbb{CP}}\!\!~^3$ $\times$ I $\times$ I}
\label{subsection:Appendix 1 A}
In this section we report the polynomials for the solution obtained via SL(2) T-duality. Using
\begin{align}
\label{eq:B.1.1}
\Delta_s & = h u'-u h', \qquad \Delta_r  = h u'+u h'
\end{align}
We write here the definition of the polynomials
\begin{align}
\label{eq NATD poly}
p_1(r) & = \frac{\pi  \left(-3 \Delta_s h u^2 h''^2 \left(\Delta_s + 2 h u'\right)+6 h^2 u^4 h''^3- \Delta_s^2 h h''' u \Delta_r -6 \Delta_s^2 h u h' h'' u'\right)}{2 \Delta^2}, \nn \\[2mm]
p_2(r) & = 2 \pi^2 \left((r-k) h' - 2h + \frac{h u \left(h'\Delta_s +\Delta_r (r-k) h''\right)}{\Delta }\right), \nn \\[2mm]
p_3(r,\rho) & = -\frac{4 \pi \rho h u'}{u}, \nn \\[2mm]
p_4(r) & = \frac{ 4 \pi^3 h u'}{3 \Delta^2} \bigg(4 h^3 h''' u^3 + 2 h^2 u^2 h'' \left( 5 \Delta_s - 2 h u'\right)+ \Delta_s^2 h \left(2 u h' + 3 \Delta_r \right) \nn \\[2mm]
& +(r-k) \left(2 h^2 u^2( \Delta_s  h''' +u h''^2 ) + 2 \Delta_s^3 h' + \Delta_s h u h'' \left(4 \Delta_s+ \Delta_r\right)\right)\bigg), \nn \\[2mm]
q(r,\rho) & = f(r)+2 \pi  \rho^2 \left(2 h' + (r-k) \left(-2 h''+(r-k) h'''\right)\right), \nn \\[2mm]
f'(r) & = \bigg( 2 u \big(6 h^2 u^3 h''^2 -h' \Delta_s^2 \left(3 \Delta_r +2 h u'\right)+2 h^2 h''' u^2 \left( \Delta_s-2 u h'\right) \nn \\[2mm]
& + h u h'' \left( \Delta_s^2+2 \left(u^2 h'^2+h^2 u'^2\right)\right)\big) -2 (r-k) \big(2 h u^3 h''^2 \left(4 h u'-9 \Delta_s\right)+8 \Delta_s^3 h' u' \nn \\[2mm]
& + 2 h h''' u^2 \left( \Delta_r^2-4 u^2 h'^2 \right) +\Delta_s u h'' \left(9 \Delta_s^2 +2 h u' \left( \Delta_r +2 u h'\right)\right)\big) \nn \\[2mm]
& -3 (r-k)^2 u \left(\Delta_r \Delta_s^2 h''' +3 \Delta_s (\Delta_r +2 \Delta_s) u h''^2-6 h u^3 h''^3+6 \Delta_s^2 h' h'' u'\right)\bigg)\frac{4 \pi^3 h}{3 \Delta ^2} 
\end{align}

\subsection{Polynomials in AdS$_2$ $\times$ $\widehat{\mathbb{CP}}\!\!~^3$ $\times$ I $\times$ S$^1$}
\label{subsection:Appendix 1 B}
In this section we report the polynomials for the solution obtained via U(1) T-duality
\begin{align}
\label{eq ATD poly}
p_5(r) & = \frac{\pi  h u \left(6 h u^3 h''^3-\Delta_s^2 h''' \Delta_r -6 \Delta_s^2 h' h'' u'- 3 h''^2  \Delta_s u \left(2 \Delta_s + \Delta_r \right)\right)}{4 \Delta ^2}, \nn \\[2mm]
p_6(r,\rho) & = \text{constant} + 2 \rho \left(h''-(r-k) h'''\right), \nn \\[2mm]
p_7(r) & = \frac{4 \pi \Delta_s u'}{u^2}, \nn \\[2mm]
p_8(r) & = -\frac{4 \pi  \left(h u' + u \left(2 h' - (r-k) \left(2 h'' - (r-k) h''' \right)\right)\right)}{u}, \nn \\[2mm]
p_9(r) & = -\frac{12 \pi  h u'}{u}, \nn \\[2mm]
p_{10}(r) & = \frac{4 \pi ^2 \Delta_s h^2 u'}{\Delta }, \nn \\[2mm]
Q'(r) & = \big(-\left(2 \Delta_s^3 h' u'+2 \Delta_s h h''' u^3 h' + u h'' \Delta_s \left(2 \Delta_s^2 +u h'\Delta_r \right) + 2 h u^3 h''^2 \left(u h'-2 \Delta_s\right)\right) \nn \\[2mm]
& +(r-k) u \left(6 h u^3 h''^3- \Delta_s^2 \Delta_r h'''-6 \Delta_s^2 h' h'' u'-3 \Delta_s u h''^2 \left(2 \Delta_s+\Delta_r \right)\right)\big)\frac{\pi^2 h}{\Delta^2} . 
\end{align}

\end{document}